\documentclass[]{interact}
\usepackage{graphicx}
\usepackage[numbers,sort&compress,merge]{natbib}
\bibpunct[, ]{[}{]}{,}{n}{,}{,}

\usepackage{color}
\usepackage{soul,xcolor}
\usepackage[normalem]{ulem}
\newcommand{\red}[1]{#1}

\begin{document}
\date{\today}

\title{Formation of Argon Cluster with Proton Seeding}

\author{
\name{O.C.F. Brown\textsuperscript{a,b}, D. Vrinceanu\textsuperscript{c}, V. Kharchenko\textsuperscript{a,d}, and H.R. Sadeghpour\textsuperscript{a}}
\affil{
\textsuperscript{a} ITAMP, Harvard-Smithsonian Center for Astrophysics, Cambridge, Massachusetts 02138, USA;
\textsuperscript{b} Department of Physics and Astronomy, University of Southampton, Southampton, SO17 1BJ, UK;
\textsuperscript{c} Department of Physics, Texas Southern University, Houston, TX 77004, USA;
\textsuperscript{d} Department of Physics, University of Connecticut, Storrs, CT 06269, USA;
}}
\maketitle

\begin{abstract}
We employ force-field molecular dynamics simulations to investigate the kinetics of nucleation to new liquid or solid phases in a dense gas of particles, seeded with ions. We use precise  atomic pair interactions, with physically correct long-range behavior, between argon atoms and protons.
Time-dependence of molecular cluster formation is analyzed at different proton concentration, temperature and argon gas density. The modified phase transitions with proton seeding of the argon gas are identified and analyzed. The seeding of the gas enhances the formation of nano-size atomic clusters and their  aggregation. The strong attraction between protons and bath gas atoms stabilizes large nano-clusters and the critical temperature for evaporation. An analytical model is proposed to describe the stability of  argon-proton  droplets, and is compared with the molecular dynamics simulations.
\\\resizebox{25pc}{!}{\includegraphics{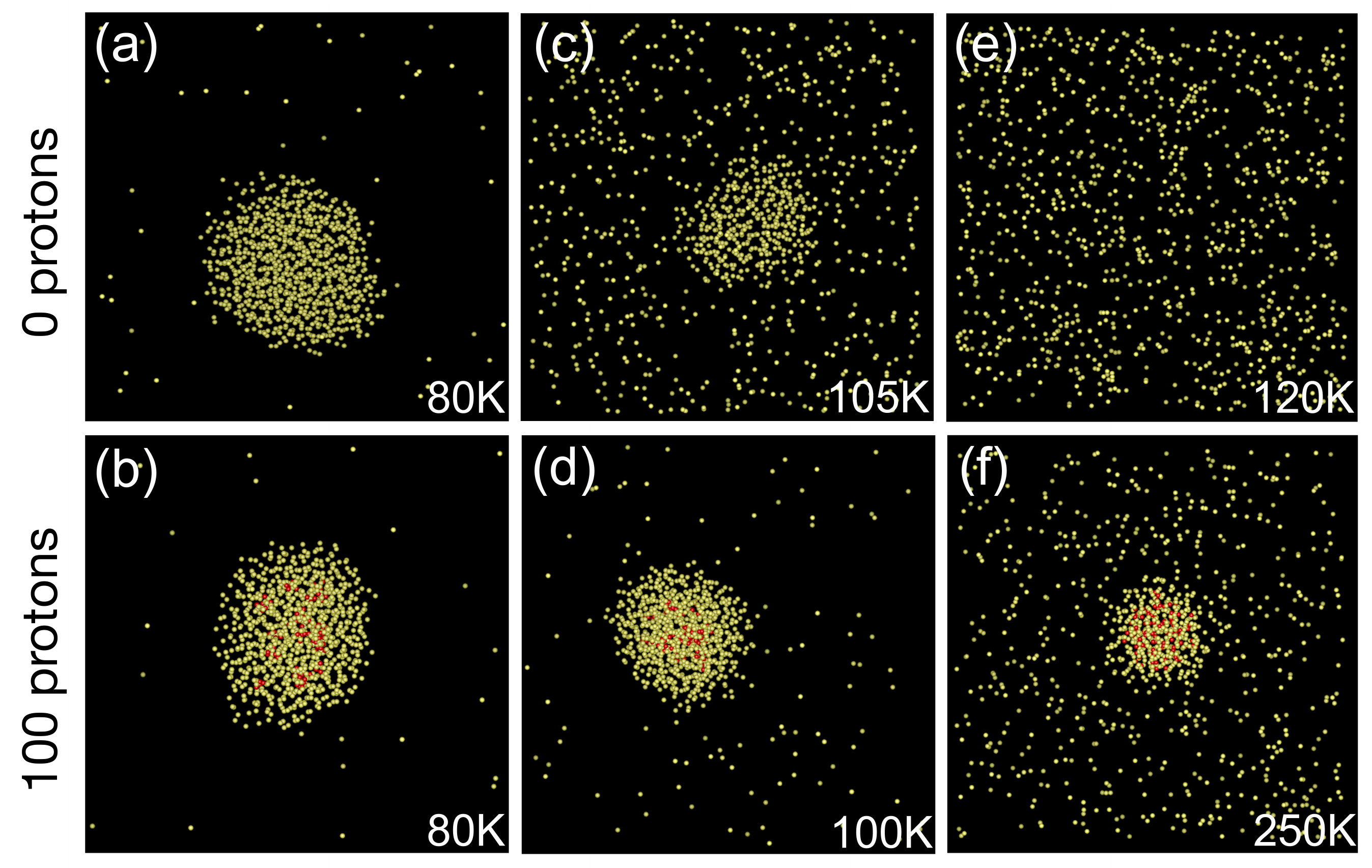}}
\end{abstract}

\begin{keywords}
Cluster formation; Molecular dynamics simulations; Protonated argon gas
\end{keywords}

\section{Introduction}
Gas phase clusters are weakly bound aggregates comprised of either atoms or molecules, and often display chemical and physical properties that are quite distinct from those of their atomic constituents or associated bulk materials \cite{Hopkins2015,Castleman1988}. To synthesize catalysts or thin films, size-selected  gas  clusters  can be delivered to substrates to obtain materials with desired individual or collective properties \cite{Alayan2004,Alayan2006}. Thermodynamics of gas phase clusters, aerosols and clouds bears on the nature of nano-particle formation in the atmosphere and in the interstellar medium \cite{Castleman1988}.

Recent interest in optical and physical  properties of gas phase clusters has been partly stimulated  by discovery of exoplanets and  analysis of absorption and emission spectra of their hazy atmospheres. Optical and infrared spectra observed from planetary and exoplanetary atmospheres, comets, and natural satellites are sometimes unusually featureless, which are attributed to the presence of atmospheric dust, ice, haze and aerosol particles \cite{Felfli2019,Zhang2017,Deming2017}. Haze is mostly formed by  small (submicron) cluster particulates that can produce a broad continuum opacity to light. Interaction between haze particles and radiating atoms or molecules can dramatically modify absorption spectra of exoplanets \cite{Vrinceanu2004,Spake2018,Felfli2019,Szudy1996, Allard1982} and used as markers for simulation of atmospheric constituents. Recent laboratory experiments simulating hazy environments for super-Earths and mini-Neptunes atmospheres suggest that some of these atmospheres contain thick photochemically generated hazes \cite{Horst2018}. 

Compositions and parameters of haze particle and atmospheric gases are expected to vary considerably for different exoplanets. The most realistic atmospheric haze materials are water and CO$_2$ ices and liquid droplets for the terrestrial atmospheres and  methane or hydrocarbons for the Jupiter-type exoplanets and Titan \cite{Tolbert2013}. Investigation of the haze formation in most important atmospheric gases, such as CO$_2$,H$_2$O, CH$_4$ etc., is a formidable task due to a complexity of molecular quantum interaction in poly-atomic gases. Cluster formation in noble gases represents an ideal environment for the laboratory simulations and theoretical modeling of spectral changes induced by  ultra-small haze particles and aggregates.

Argon is a potential target species in search for naturally occurring, noble gas compounds. Argon is known to be polarizable with the proper ligands \cite{Pauzat2013}. While argon clusters are some of the simplest chemical systems to study,  experimental investigations of the structure and stability of neutral rare-gas clusters {and nano-size liquid droplets} are  extremely challenging. Charged clusters are on the other hand, easily studied using mass spectrometric techniques \cite{Gatchell2018,Shen2015}. The first noble gas molecule observed in nature is the simple argon-proton cation (argonium, Ar-p) \cite{Cody2017}. Argonium has been detected in the interstellar medium (ISM) toward various astronomical objects \cite{Fortenberry2017,Schilke2014}. Also, the existence of the proton-bound dimer Ar-p-Ar has been proved by spectroscopic evidence in argon matrix \cite{Bondybey1972,Kunttu1994}. Ion-molecule complexes of the form Ar$_n$p are detected in pulsed-discharge supersonic expansions containing hydrogen and argon \cite{McDonald2016}, and studied theoretically \cite{Giju2002}. The nucleation stimulated by proton seeding is unique because proton has no core electrons, and from the chemical point of view it  can be seen as a  point charge \cite{Beyer1999}.

Considerable effort has been devoted to obtaining a better understanding of the nucleation \cite{Kraska2006,Laasonen2000}, structural properties \cite{Hoang2008}, collision dynamics \cite{Ming1997}, and phase transitions \cite{Dumont1995,Rey1992} of argon clusters by means of classical Molecular Dynamics (MD) \cite{Baletto2005}. The dependence of argon phase transformations  on the size of clusters is investigated experimentally by using electron diffraction analysis \cite{Danylchenko2004, Danilachenko2008,Danylchenko2014}, and  was  predicted theoretically using MD simulations \cite{Bingxi2017}. Phase diagram of argon nano-clusters up to 400 atoms has been reported by means of constant energy molecular dynamics simulations \cite{Manninen1998}.

The cluster nucleation in atomic and molecular gases occurs in several stages. Phase transitions  from a thermodynamically metastable state to a stable state  occur in the homogeneous gas due to microscopic fluctuations.  Fluctuations produce  nano-size clusters in a liquid or solid  phase. These clusters are relatively stable and become centers of growth of a new phase, if the  typical cluster size $R$ is  larger than some critical value $R_c$. Clusters with $R<R_c$ are unstable and disappear back into the gas phase \cite{Pitaevskii}. Ions and other seed particles may stimulate formation of critical clusters arising in the early stage of nucleation. Modeling  of production of critical clusters is the  most  difficult part in investigations of nucleation processes. We performed MD simulations aimed at clarifying the kinetics of  short-term nucleation, which initiates sub-critical and critical clusters. The long-term stages of haze formation, such as coalescence, when  growth of larger clusters occurs due to "swallowing up" of small ones \cite{Pitaevskii}, can then described with standard kinetics of the first order phase transition. 

Charge particles can catalyze short-term cluster formation in the gas phase. A main goal of this work is to study argon nucleation with and without proton seeding, based on the most accurate quantum-mechanical binary potentials for classical MD simulations. We will show how small concentrations of positive ions accelerate nucleation process, but high ionic densities prevent formation of the gas phase clusters. The phase transitions in clusters due to the temperature change and proton contribution are studied by analyzing the pair correlation functions (the radial distribution function - RDF), and the size of nucleation clusters. \red{ Other order parameters, or discriminating quantities such as  mean square displacement, or diffusion coefficient, are more adapted to studying bulk transformations because their dependence on cluster size makes these parameters less unique. In future extensions of this work for larger clusters and aggregates, these complementary measures will be investigated.}

\section{Simulation procedure and details}
The Large Atomic/Molecular Massively Parallel Simulator (LAMMPS) \cite{LAMMPS1995} is employed to perform simulations of the dynamics of cluster formation in an Ar gas when seeded with protons. The classical force fields are calculated from the quantum mechanical pair interaction potentials which are described in detail below. The results are visualized using Visual Molecular Dynamics software (VMD)\cite{VMD1996}. 

\subsection{Binary interaction potentials}
The dynamics of cluster formation, stability, and structural properties with ion seeding are obtained from MD simulations that use classical force fields deduced from accurate quantum calculation of pairwise interaction potentials.

{ \it Ar-Ar interaction.} The binary Ar-Ar potential is modeled by a Lennard-Jones 6-12 (LJ) potential \cite{Manninen1998,White1999,Henderson1969}. 
\begin{equation}\label{VArAr}
	V_{Ar-Ar}(r) =4\varepsilon \big[(\frac{\sigma}{r})^{12} - (\frac{\sigma}{r})^6], 
\end{equation}
where parameters, $\varepsilon =1.23\times 10^{-2}$ eV and $\sigma =3.357$ \AA, are deduced from gas and matrix spectroscopy. The physically correct weak van der Waals (vdW) asymptotic behavior is evident from the LJ potential.

{\it Ar-p interaction.} The Ar-p Born-Oppenheimer (BO) potential energy curves for the ground and excited electronic states were calculated by Sidis \cite{Sidis1972}. The BO potentials were subsequently used for the calculations of the scattering phase shifts and compared with existing experimental data, which in turn used to improve the well depth of the {\it ab initio} results.

The long-range form of the Ar-p energy is dominated by the polarization potential Eq.~\ref{ArP_III}, due to the polarizability of the Ar ground state. This long-range potential (region III) is connected to the BO potential at intermediate and short-range (region I), by a switching function (region II).

At short distances, the Ar-p energy is modeled by a Morse-type potential as \cite{Sidis1972},
\begin{equation}\label{ArP_I}
	V_{Ar-p}^{(I)}(r) =U\times(x^{2}-2x),
\end{equation}
with
\begin{equation}\label{ArP_x}
	x=(\frac{r_{e}}{r})^{a}e^{[b(r_{e}-r)]}
\end{equation}
where $U=4.3$ eV, $r_{e}=1.34$ \AA, and,
 \[
    \left\{
                \begin{array}{ll}
                  a=0.475,  b=1.370\mbox{ \AA}^{-1} \quad \textrm{when} \quad   r < 1.34\mbox{ \AA} \\
                  a=0.400,  b=1.890\mbox{ \AA}^{-1}  \quad \textrm{when} \quad  r > 1.34\mbox{ \AA}
                \end{array}
              \right.
  \]\\
The switching function is defined in the interval [3.6 \AA, 3.86 \AA] (region II) as,
\begin{equation}\label{ArP_II}
V_{Ar-p}^{(II)} (r) =\frac{A}{e^{\gamma (r-r_{s})}+1}+B\times r+C.
\end{equation}
where $A=-0.095$ eV, $B=0.01$ eV/\AA, $C=-0.085$ eV, $\gamma=7.0$ \AA$^{-1}$, and $r_{s}=3.5$ \AA.
The Ar-p interaction potential at long range is given by,
\begin{equation}\label{ArP_III}
V_{Ar-p}^{(III)} (r) = - \frac{q_p^2 \alpha}{ 8 \pi \varepsilon_0 r^{4}}.
\end{equation}
where $\alpha=11.08$ $a_{0}^3$ is the argon static polarizability in the ground state, $q_p$ is the ionic charge in a charge neutral plasma, $\varepsilon_0$ is the permittivity of free space, and $a_0$ is the atomic unit of length.

{\it p-p interaction.} The Coulomb repulsion between two protons is shielded in a neutral plasma by the free electrons. This is modeled by Debye shielding feature available in LAMMPS. The Debye length is \cite{Jackson1975},
\begin{equation}\label{Debye}
\lambda_D = \sqrt{\frac{\varepsilon_0k_B}{q_p^2\left(\frac{\rho_e}{T_e} + \frac{\rho_{p}}{T_p}\right)}},
\end{equation}
where $k_B$ is the Boltzmann constant, $T_p$ and $T_e$ are the temperatures of the ions and electrons respectively, $\rho_{p}$ and $\rho_e$ are the ion and electron number densities,  respectively. The screened potential is,
\begin{equation}\label{Coulomb}
V_{p-p}(r) = \frac{q_p^2}{4\pi\varepsilon_0r}e^{- r/\lambda_D}.
\end{equation}

\begin{figure}
\includegraphics[width=1.0\linewidth]{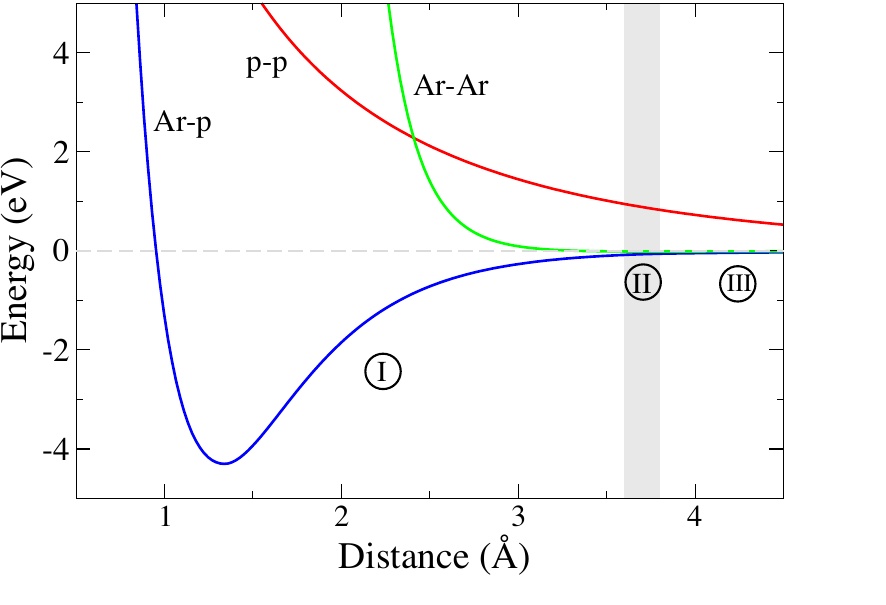}
\caption{\label{fig:potentials} Potential energy curves of pair interactions for the MD simulations. The Ar-Ar interaction (green) is modeled by LJ potential, Eq.~(1). The screened proton-proton potential Eq.~(\ref{Debye}) (red) is calculated at a particular density with $\lambda_{D}=2.5$ \AA~using the Debye shielding feature available in LAMMPS. The potential energy between argon atoms and protons (blue) is defined in short (I), and long (III) ranges, with a switching function connecting them in region (II), Eqs.~(\ref{ArP_I})-(\ref{ArP_III}).}
\end{figure}

The interaction potentials are displayed in Fig.~\ref{fig:potentials}. While the Ar-Ar BO potential contains a shallow well in the region II and is asymptotically attractive due to long range vdW forces, the greatest attraction in the simulations is understandably due to the Ar-p potential in region I. The competition between the screened p-p and Ar-p interactions ultimately determines the formation and stability of the clusters as a function of proton concentration.
 
\subsection{MD simulations}
Each MD simulation starts by generating random coordinates for a given number of atoms and ions, with a minimum separation of 3 \AA. Simulations are performed with a fixed number of Ar atoms (1000) and a variable number of protons, from 0 to 100, while adjusting the simulation cell in order to obtain various densities. The simulations are performed with the canonical (NVT) ensembles with a Nos\'e-Hoover thermostat \cite{LAMMPS1995}.

Since the Debye length depends on  temperature, simulations are run at various temperatures with adjusted Debye length. Charge neutrality in the simulation cell is preserved by specifying an equal number of electrons and protons in Eq.(\ref{Debye}).

We are interested in the short-time dynamics of Ar and Ar-p cluster nucleations. Experimentally, the proton seeding could be implemented by selective laser ionization of H atoms in the Ar and H gas mixture. Energies of  photo-electrons can be significantly larger than  the bath gas temperature. Energy relaxation of electrons occurs mostly in collision with the Ar atoms and the electron plasma thermalization requires significantly more time than the characteristic time of cluster formation. For example, electrons with energies around 11.5 eV can form the long-living metastable ions Ar$^{-}$(3p$^5$4s4p) with the lifetime $\sim$ 260-300 ns \cite{NegativeIons}. This time is much larger than a time of $\sim$10 ns required for the production of critical clusters.  Energetic electrons do not play a significant role in Debye screening and, aside from providing the plasma neutrality, they are not material to the simulations.
 
The average time between collision of atomic particles  $\tau_{col}=1/(\rho_{Ar}\sigma_{Ar}\langle{v}\rangle)$  has to be significantly larger than the simulation time step, $\delta t$. Simple but realistic assumptions for the radius of Ar atoms $r_{Ar} \sim 10^{-8}$ cm and for the elastic cross section of Ar-Ar collisions , $\sigma_{Ar}\sim\pi  r_{Ar}^2  $ allow us to estimate $\tau_{col}$ at the  argon number density of $\rho_{Ar}=10^{21}$ cm$^{-3}$, and an average velocity $\langle{v}\rangle \sim 10^{4}$ cm s$^{-1}$, as  $\tau_{col} \sim  0.3$ ns. We therefore use a fine time-step of $\delta t$=1.0 fs $\ll \tau_{col}$ for all the simulations here. {For our simulations, this time resolution is always less than the collision time and therefore gives us the capability to observe the dynamics at high densities.} To determine the number of atoms in a cluster, the distance between atoms were evaluated during the simulation and any that were within 6 \AA~of each other were counted to be part of a single cluster. This distance was chosen based on the lattice constant of the Ar crystal, $a=5.26$\AA~\cite{Hermann2011}. Dispersed small clusters eventually aggregate to form one large cluster at longer times.

\section{Results and Discussion}
\subsection{Ar solid phase transition without proton seeding}
We first study the crystallization of argon at a high density, without proton seeding. This process of the cluster formation is an excellent example of homogeneous nucleation \cite{Pitaevskii}. The simulations are performed at high  densities of Ar atoms $\rho_{Ar} \sim 10^{22}$ cm$^{-3}$ and T=40 K. This density is achieved by positioning 1000 argon atoms regularly in a cube at 4.62 \AA\ separation. At this density, $\langle{r}\rangle \sim \rho_{Ar}^{-1/3}\sim10$\AA~$ > \sigma$. At such mean separations, the Ar atoms are ensured on average, to experience each other through the attractive vdW interactions. It can therefore be predicted that the system quickly transforms into a solid phase, which can  have  an amorphous or polycrystalline structure.

\begin{figure}
\includegraphics[width=1.0\linewidth]{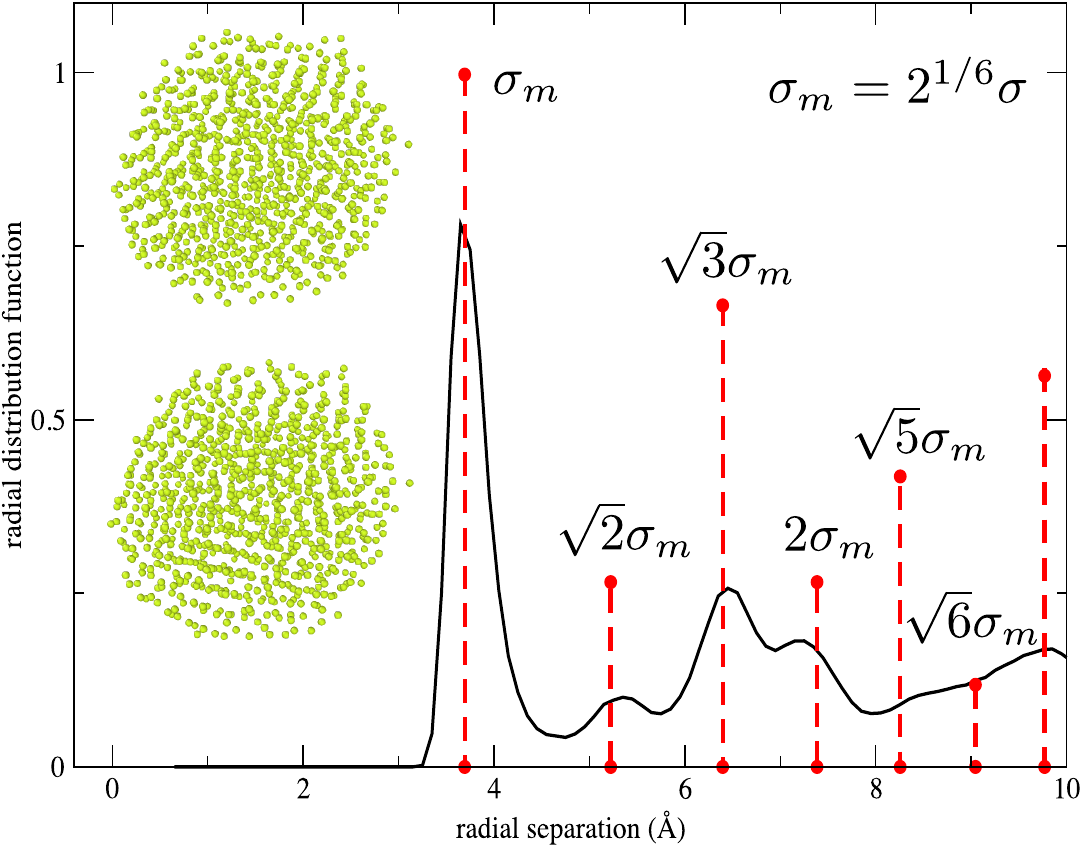}
\caption{\label{fig:crystalline} 
\red{ The Ar-Ar RDF of pure solid argon cluster formed from initial density of 10$^{22}$ cm$^{-3}$ after 10 ns of MD simulation at 40 K.  The inset shows snapshots of atom configuration from two different view points. The bottom cluster viewpoint in the inset shows more clearly the lattice structure. The vertical dashed lines indicate the coordination spheres of the ideal FCC crystal labeled by irrational multiples of the LJ parameter $\sigma =3.357$ \AA. The height of the vertical lines are the coordination numbers scaled by the square of the corresponding radius of coordination sphere, and normalized to unity for the $\sigma_m$ peak.}}
\end{figure}

The inset in Fig.~\ref{fig:crystalline} shows the final configuration of this system after 10 ns of MD simulation. This {crystalline and/or amorphous} structure is compared with the perfect fcc argon crystal that has atoms separated by precise distances. A meaningful measure of phase transition is the radial distribution functions (RDF), presented in Figs.~\ref{fig:crystalline} and computed as,  
\begin{equation}\label{eqn_1}
g_{ab}=\frac{\langle{\rho_{b}(r)}\rangle}{\langle{\rho_{b}}\rangle_{loc}}
\end{equation}
where $\langle{\rho_{b}(r)}\rangle$ refers to the average density of particle $b$ at a distance $r$ from particle $a$, and $\langle{\rho_{b}}\rangle_{loc}$ refers to the density of the particles $b$, as if they are uniformly distributed within the simulation cell.

\red{ Fig.~\ref{fig:crystalline} demonstrates the formation of an amorphous/crystalline structure during the simulation. The peaks represent particle densities distributed around their respective bulk lattice positions indicated by the vertical dashed lines labeled by the irrational multiples ($\sqrt{2}$, $\sqrt{3}$, $\ldots$) of the inter Ar atom equilibrium separation parameter, $\sigma_m = 2^{1/6} \sigma$ \cite{Chandler1987}.}

\red{ The height of the vertical bars are proportional to the coordination numbers corresponding to the ideal crystal positions, and inversely proportional to the square of their distances. The shape of the RDF curve follows roughly the same pattern, with maxima around lattice separations. The agreement is qualitative and is affected by thermal motion, finite size of the sample and large surface effects, explaining for example why the RDF is non-zero in between ideal crystal sites. The spherical-like shape of final configuration depends on the way of building the starting configuration, although the RDF curves are stable, and only depend on the temperature and the number of particles.}

\subsection{Phase transition with proton seeding}
Argon in the gas phase is normally {mono-atomic} \cite{Christe2001}. The  weak inter-atomic interactions between  Ar atoms is manifested in Ar low {normal} melting $T_{m}\simeq83.8 K$ \cite{Zelfde1968} and boiling  $T_{b}=87.28 K$  \cite{Grossman1969} temperatures. This behavior can be predicted from an analysis of Ar-Ar inter-atomic potential that has a shallow attractive well and a steep repulsive wall, as shown in Fig.~(\ref{fig:potentials}). 

\begin{figure}
\includegraphics[width=1.0\linewidth]{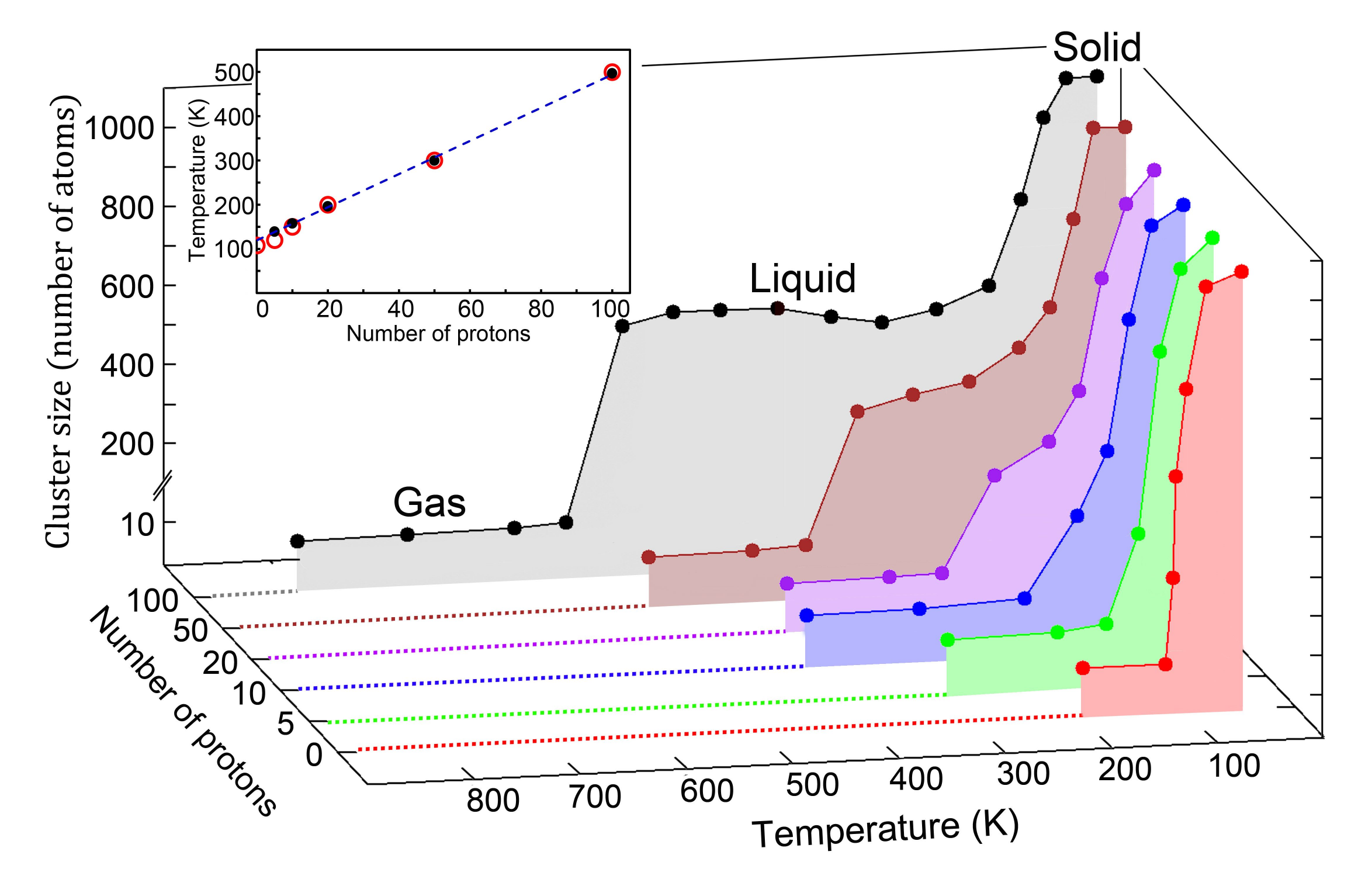}
\caption{\label{fig:growing1} The average size of clusters plotted against temperature with various number of protons,$N_{p}$, when the initial argon density is 10$^{21}$ cm$^{-3}$. The inset shows the highest temperature, $T_{crit}$, at which stable clusters still exist, for each of the simulated systems. The red empty circles are from the MD simulations, and the solid black circles are predictions from Eq.~\ref{tcrit}.}
\end{figure}

\begin{figure}
\includegraphics[width=1.0\linewidth]{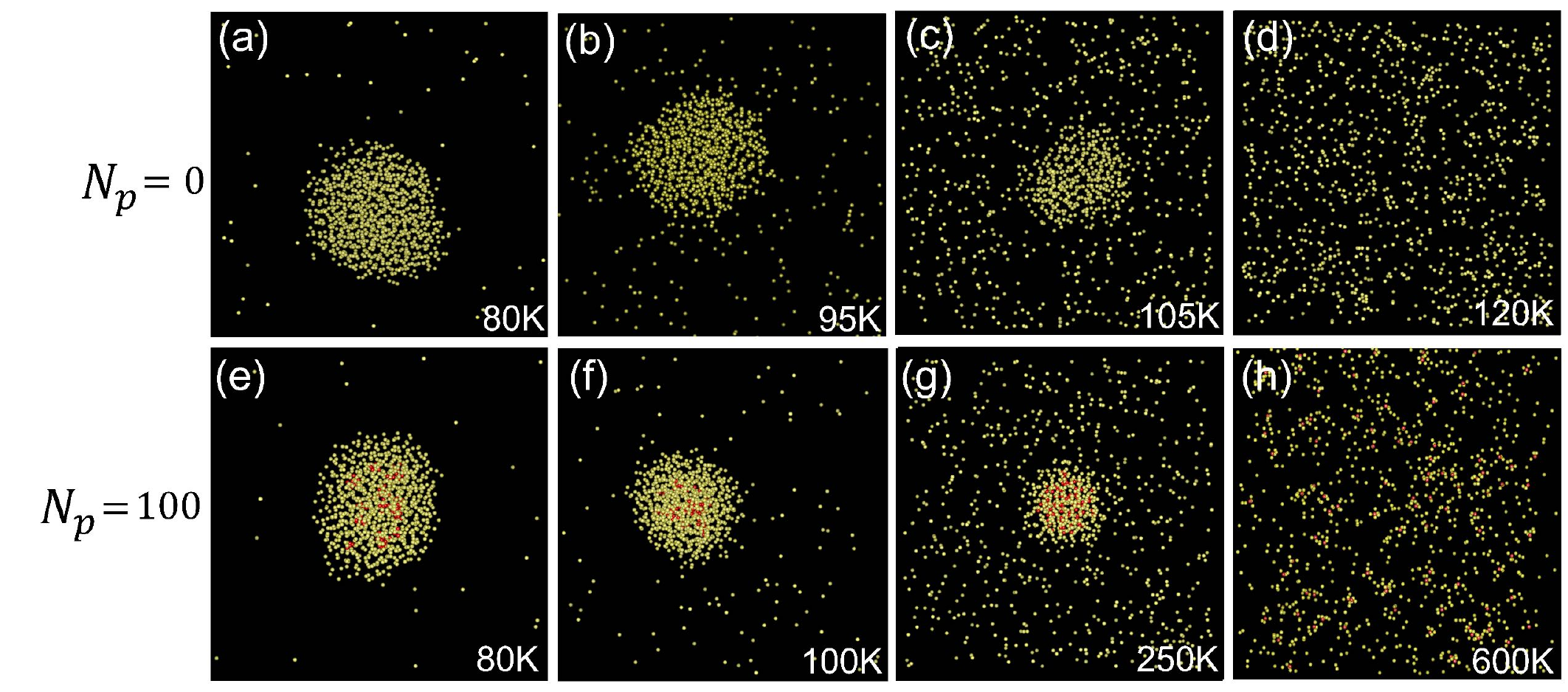}
\caption{\label{fig:snapshots} Snapshots of clusters formed without protons (top panels, a-d) and with 100 protons (bottom panels, e-h). The argon density is 10$^{21}$ cm$^{-3}$. Particles coloured in red are protons, whilst yellow particles are argon atoms. Up to the critical temperature, $T_{crit}$, less than $10\%$ of protons are found outside the clusters.}
\end{figure}

To examine the phase transition of argon, simulations are performed at $10^{21}$ cm$^{-3}$ argon density in a simulation cell with length of 100 \AA~at T=50 K. Phase diagram of pure argon \cite{Bolmatov2015} indicates that the system at this point is still in the solid phase. The simulation is run for 10 ns to ensure that a single solid phase cluster of all the atoms has formed at the steady state condition. The size of clusters are then monitored as the temperature gradually increases until clusters no longer exist. The results are shown in Fig.~(\ref{fig:growing1}). The red curve, for instance, in the absence of protons, shows temperatures up to T=200 K, and a transition from the solid to the gas phase at a temperature between T=80 K to 120 K. Transitions from  the crystalline solid state to gas phase can occur via intermediate amorphous or liquid cluster states. These phases  have been detected by analyzing the radial pair correlation functions. This can also be seen clearly from the snapshots of the system given in the top panels in Fig.~\ref{fig:snapshots}(a-d).

The proton seeding effect on the cluster formation is investigated by  increasing the proton density in the MD simulation box. The results are illustrated in Fig.~(\ref{fig:growing1}) when $N_{p}=$ 5, 10, 20, 50, 100 protons are randomly added to the simulation cell which already has 1000 randomly positioned argon atoms. 

\begin{figure}
\includegraphics[width=1.0\linewidth]{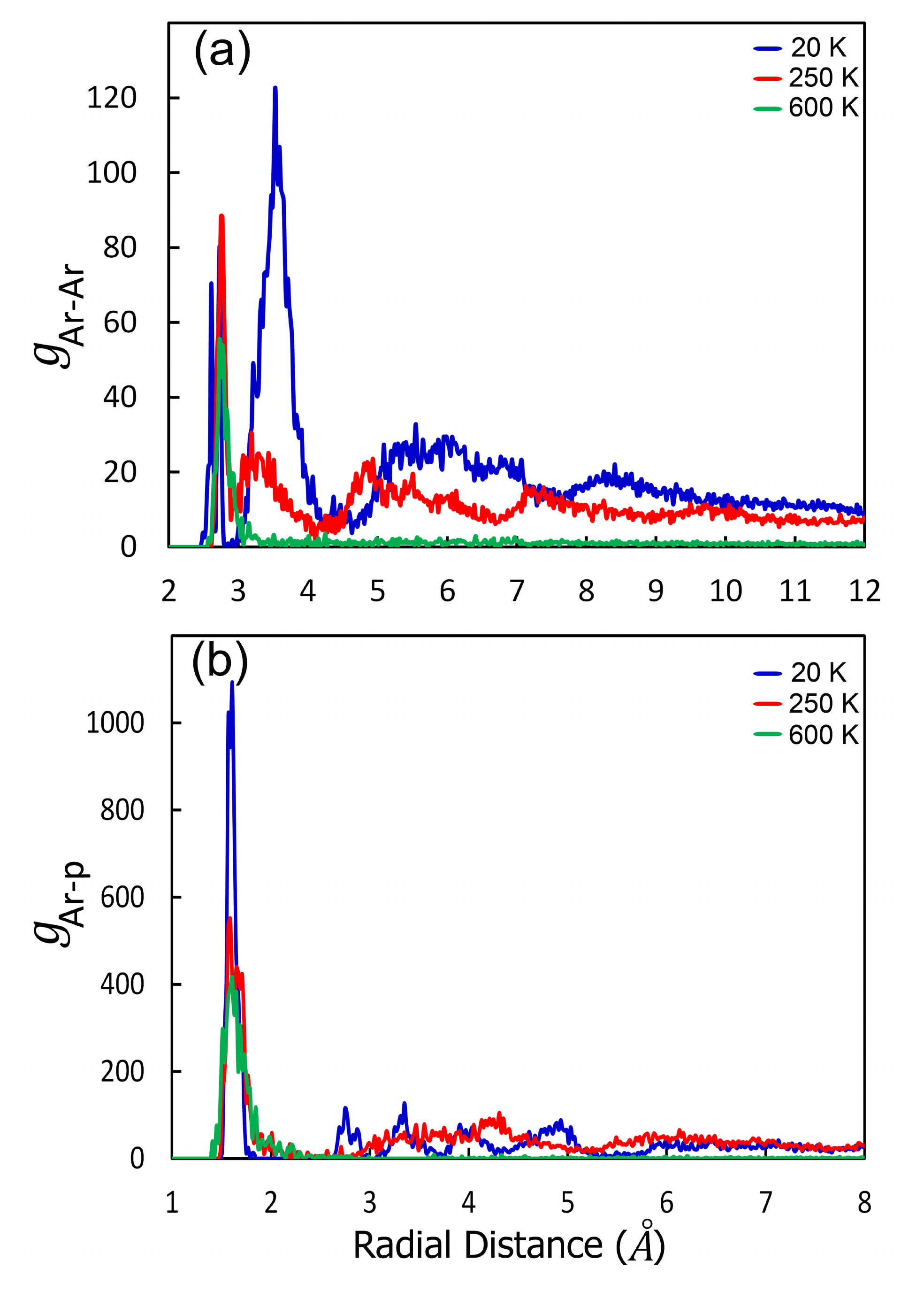}
\caption{\label{fig:RDF} The pair-correlation (RDF) curves for (a) Ar-Ar, and (b) Ar-p distributions at various temperatures at argon density 10$^{21}$ cm$^{-3}$, and proton density  10$^{20}$ cm$^{-3}$. The characteristic peaks for the solid, liquid, and gas phases are distinctly visible at T=20, 250, and 600 K, respectively. \red{ In addition to RDF, we also use the size of the cluster (the number of Ar atoms in the cluster) to distinguish the appearance of different phases.}}
\end{figure}

The most prominent feature in Fig.~(\ref{fig:growing1}) is the appearance of a distinct plateau with proton seeding. This plateau indicates the formation of a stable liquid phase. The reader is referred to the pair-correlation functions in Fig.~\ref{fig:RDF} for a clear demonstration of the liquid phase plateau. The phase modification with proton seeding is observed as the liquid-Ar plateau widens with increasing proton number, and the solid phase melting point moves to higher temperatures with increasing proton number. The effectiveness of the seeding process on cluster formation and its stability can also be seen in the inset of Fig.~(\ref{fig:growing1}). As will be discussed in more details later (in Sec. D), this figure shows that the proton mediated nucleation and clustering makes the liquid-gas transition to linearly shift to higher temperatures with respect to $N_{p}$. Relatively strong attraction between protons and Ar atoms, see Fig.~\ref{fig:potentials}, creates a favorable conditions for the Ar clustering and protons work as a strong "ionic glue" for Ar atoms. Nevertheless, one should expect that at very high proton densities Coulomb's repulsion  between ions will stimulate cluster decay.

The snapshots in the bottom panels in Fig.~\ref{fig:snapshots}(e,f,g) visualize the formation of the droplets at a proton density of $\rho_{p}=10^{20}$ cm$^{-3}$ at higher temperatures. The transition to the gas phase in this case happens at $T>500$ K, Fig.~\ref{fig:snapshots}(h). 
As can be seen from the plateau in Fig.~(\ref{fig:growing1}), the size of clusters in the liquid phase, at temperatures T=200 to 500 K, stays roughly constant, \textit{i.e.} around 500 argon atoms. By measuring the volume of these clusters, the Ar number density of the droplets are calculated to be $\sim 1.2\times 10^{21}$ cm$^{-3}$. 

\subsection{Radial distribution function analysis}
The pair-correlation radial distribution function (RDF) analysis is performed to confirm the nature of the phases at different temperatures. Figure~\ref{fig:RDF}(a) shows the Ar-Ar RDF curves for at T=20 K (solid), T=250 K (liquid), and T=600 K (gas). The curves exhibit characteristic signatures  for corresponding phases with temperature \cite{Rossler2009}. The relatively periodic and discrete peaks in the blue curve at T=20 K demonstrate the coordination shell for the solid structures (also indicated in Fig.~\ref{fig:crystalline}(b)). The red curve at T=250 K illustrates a similar sharp peak, followed by a few smaller subsequent peaks. The absence of exact intervals in between these peaks is an indication of a loosely packed structure, which is a known feature in liquids. Due to increased mobility, atoms in the liquid state do not maintain a constant structure. Since particles become independent of each other at large distances, their RDF do not maintain long range order, and the distribution returns to the bulk density. Finally, the single peak in the green curve at T=600 K with a rapidly decaying form is the characteristic feature of the gas phase. Similar features appear in Fig.~\ref{fig:RDF}(b) for the Ar-p RDF curves. 

Visualization of the final configurations in the gas phase shows the existence of a tetrahedral-like structure with a proton at the center. This structure which is relatively stable up to the highest temperatures we simulated, \textit{i.e.} T=800K, may be an indication for the coordination number $\sim 4$ for the p-Ar system.

\red{ The behavior of the RDF with temperature, combined with the cluster size (number of Ar atoms in the cluster) pattern in Fig. 3, confirm how the long range order changes in the cluster with increasing $T$, leading to the transition through different phases, when seeded with protons.}

\subsection{{Critical temperature  for protonated clusters}}

The proton interaction with Ar atoms and  proton-proton  correlation significantly contribute to the formation of stable  proton-argon clusters. The charge screening, a specific type of correlation effects, reduces  the total energy of the Ar-p clusters, and makes them  more stable. We consider  a neutral proton and electron plasma in the Ar bath gas. The Debye screening length, $\lambda_D$, depends mostly on the proton density and temperature, 
\begin{equation}
\lambda_D \simeq 
\sqrt{\frac{\varepsilon_0 k_B T_p }{q_p^2  \rho_{p}}} ~  \Leftrightarrow ~  k_B T_p = 4\pi \left( {q_p^2\over 4\pi \varepsilon_0 \lambda_D} \right) \rho_p \lambda_D^3.
\end{equation}
where $T_{p}$, $q_{p}$, $\rho_{p}$ are the temperature, charge, and the number density of protons in the simulation cell, respectively.

The efficiency of the Debye screening depends on the number of free charged particles $n_p
\sim \rho_p \lambda_D^3 $ inside a sphere with the Debye length $\lambda_D$. The characteristic energy of the screened Coulomb interaction $ q_p^2/ 4\pi \varepsilon_0 \lambda_D$ can be used to estimate the correlation energy $\varepsilon_D= \xi~ q_p^2/ 4\pi \varepsilon_0 \lambda_D$ of the charged particles interaction. The  correction parameter $\xi(\rho_p,\rho_{Ar},T_p)$ depends on the configuration of Ar-p clusters and on the  relative strength of Ar-p, Ar-Ar and p-p interactions. $\xi$-values may be slightly different for the liquid, amorphous, or crystalline structures. For systems with relatively large proton density, the correction parameter $\xi\sim ~1$, but at low proton densities $\xi$-coefficient  may be an order of magnitude larger. The physical reason for this is a relatively strong Ar-p attraction, compared to the screened p-p repulsion at small values of $\rho_p$. 

At low charge densities the Coulomb correlation energy depends rather on the cluster radius ($\varepsilon_D \sim q_p^2/ 4\pi \varepsilon_0 R_p$), as well as the number of protons in the cluster ($n_p \sim \rho_p R_p^3$). Our MD simulations show that the radius of the Ar-p critical clusters varies in a relatively narrow interval $R_p = 11-15$ \AA~even when the proton density is altered in a broad interval of $\rho_p \sim 10^{18}-10^{20} $ cm$^{-3}$. In the numerical MD simulations of the cluster stability, the density of free protons is  $\rho_p = N_{p}/L^3 $, when the simulation box length is $L=100$ \AA.

The correlation energies are especially important for the critical states of liquid/amorphous  Ar-p clusters, where a relative small increase in temperature,or screening length, leads to the cluster evaporation. The simplified  relationship between the critical temperature, $T_{crit}$ (the highest temperature at which stable clusters still exist), and the critical proton densities, $\rho_{p_c}$, can be formulated using Eq.~(3). The fragmentation of  Ar-p clusters may  occur, if the average  thermal energy $ {3  \over2} k_B T_{crit}$ is larger than the proton correlation energy in the cluster $n_p \varepsilon_D$ or larger than the binding energies induced by Ar-p and Ar-Ar interactions. The model equation for the simplified relationship  between $T_{crit}$ and $n_{p}$ in our analysis of the Ar-p clusters can be written as,
\begin{equation}
  T_{crit}= {8 \pi\over 3} \xi \left( {q_p^2\over 4\pi \varepsilon_0 \lambda_D  k_B}\right) \times
 \left({\lambda_D \over L}\right )^3 N_{p} + T^{Ar}_{crit},
 \label{tcrit}
 \end{equation}
where $T^{Ar}_{crit}\simeq $108 K is the  critical temperature for Ar-clusters in the pure Ar gas obtained in our numerical simulations. Numerical values $\xi(N_{p}, T_{crit})$ of the correction coefficient, 
\begin{equation}
 \xi= \xi(n_{p},T_{crit}) \simeq 0.12 {R_p(n_{p},T_{crit})\over \lambda_D(n_{p},T_{crit})}
  \label{coeffs}
\end{equation}
have been computed using $R_p$-data from the MD simulations of Ar-p cluster formation.  Eq.~\ref{tcrit} can be considered as a transcendental equation for self-consistent calculations of the critical temperatures, $T_{crit}$. Comparisons between critical temperatures calculated from the analytical formula,
Eq.~\ref{tcrit}, and computed in our MD simulations are shown in the inset of Fig.~(\ref{fig:growing1}). Agreement is excellent for high proton densities. Nevertheless, for a small number of protons  the results of MD simulations and prediction of analytical formula differ about  $\sim10-15\%$ because  averaged Ar-p and Ar-Ar interactions dominate over an average value of p-p repulsive energy.

\subsection{{Clustering at lower density}}

\begin{figure}
\includegraphics[width=1.0\linewidth]{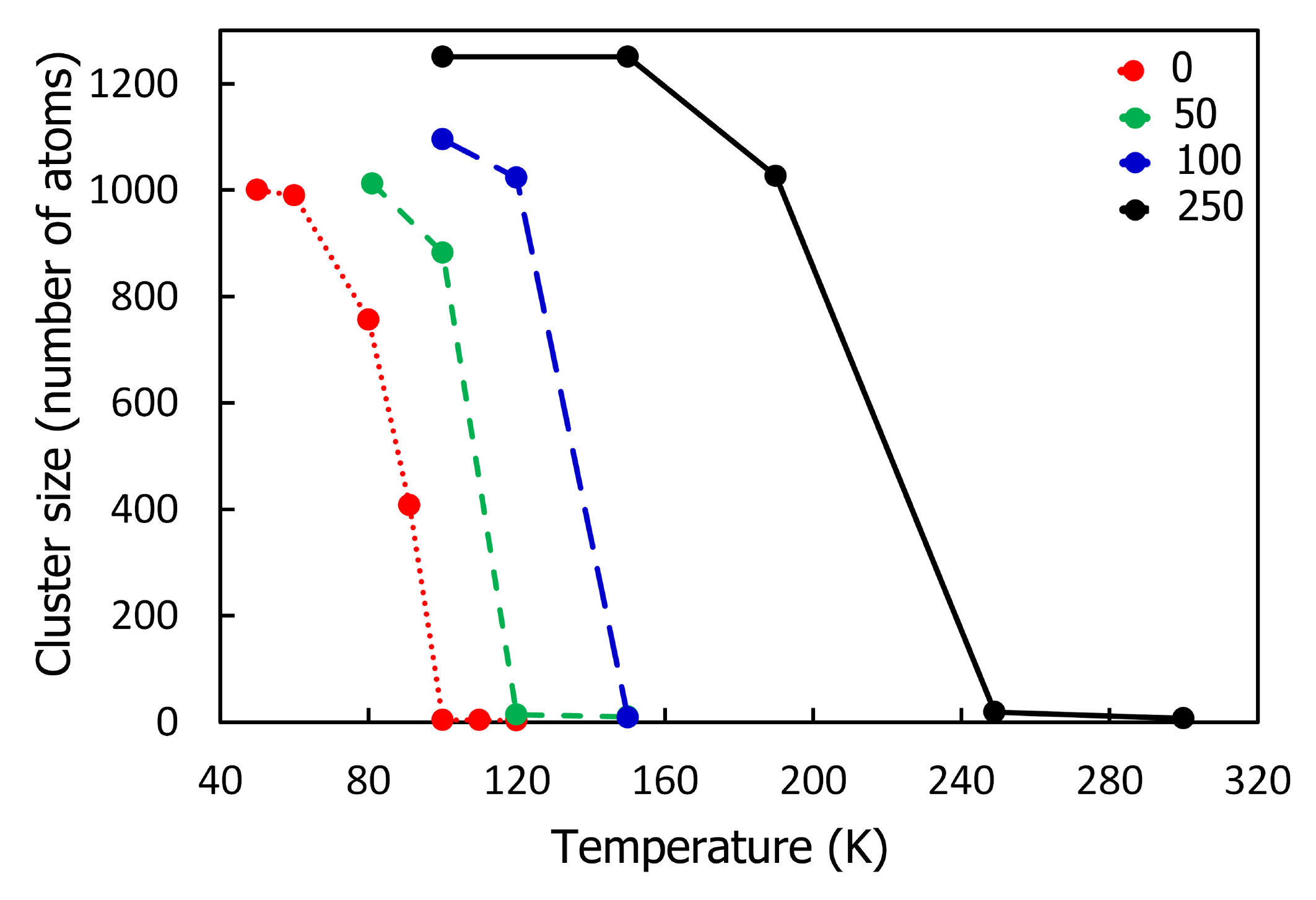}
\caption{\label{fig:growing2} The size of Ar clusters against temperature over a range of proton densities, when the initial argon density is 10$^{20}$ cm$^{-3}$ for an increasing number of protons: 0, 50, 100, and 250.}
\end{figure}

Finally, we performed simulations at a lower argon density, 10$^{20}$ cm$^{-3}$. This density is generated by increasing the size of the square simulation box to the length of 215.44 \AA, while the number of argon atoms are kept constant at 1000. Figure~(\ref{fig:growing2}) shows the change of the average cluster size with respect to temperature for different proton contributions. As can be seen in this figure, transition to the gas phase for the case of pure argon (red dotted curve) occurs at around T=100 K, compared to T=120 K when the density of argon was 10$^{21}$ cm$^{-3}$. We have found that  adding protons in the lower density Ar gas  pushes the transition to the gas phase to higher temperatures.   At the same time, no liquid phase plateau forms even at the highest number of protons we simulated, \textit{i.e.} $N_{p}=250$. This is because  the Debye length increases at low densities to become greater than the distance 
of the Ar-p interaction where is the most attractive.

\section{Summary and outlook}
Extensive molecular dynamics simulations were employed to understand the formation of the critical argon nano-clusters under different conditions. We showed that the short-term nucleation and aggregate formation are enhanced when the Ar bath gas  are seeded with proton. The effectiveness of the seeding mechanism was evaluated at different argon densities and temperatures. In particular, we observed the formation of stable liquid droplets over a wide range of (up to around T=500 K) when a 10$^{21}$ cm$^{-3}$ argon sample is protonated. The phase transition of such proton mediated argon clusters was analyzed by studying the pair-correlation (radial distribution functions) at different temperatures to confirm the solid, liquid and gaseous nature of each phase. The comparison of constant pressure phase transition lines suggest a linear increase of the temperature at which the clusters dissolve into the gas phase with increasing proton density. This indicates the efficiency of the proton seeding process to enhance the stability of the formed critical nano-clusters at higher temperatures.

Further work will attempt to improve the model beyond considering purely binary interactions between particles. We plan to include few-body molecular potentials to more accurately depict the long range attraction between an argon and a proton-argon cluster containing a few argon atoms. This will remove the long range dependence on the angle of approach of an argon atom towards a cluster. Modelling of  haze  and dust nano-particle  formation  in  astrophysical and atmospheric environments from  water vapour, methane, or other organic molecules may be considered as a next logical step of investigations. 

\section*{Acknowledgments}
OCFB was supported through a collaborative program between Southampton University and Harvard-Smithsonian Center for Astrophysics program. HRS acknowledges support from the NSF through a grant for ITAMP at Harvard  University. One of the authors (DV) is also grateful for the support received from the National Science Foundation through grants PHY-1831977 and HRD-1829184.


\end{document}